\documentclass[11pt]{article}

\usepackage[margin=1in]{geometry}
\usepackage[T1]{fontenc}
\usepackage{lmodern}
\usepackage{microtype}
\usepackage{amsmath,amssymb}
\usepackage{booktabs}
\usepackage{multirow}
\usepackage{float}
\usepackage{graphicx}
\usepackage[dvipsnames]{xcolor}
\usepackage{caption}
\usepackage[numbers,sort&compress]{natbib}
\usepackage[colorlinks=true,linkcolor=MidnightBlue,citecolor=MidnightBlue,urlcolor=MidnightBlue]{hyperref}
\usepackage{titlesec}
\usepackage{abstract}

\graphicspath{{./}}
\titleformat{\section}{\large\bfseries\color{MidnightBlue}}{\thesection}{0.6em}{}
\titleformat{\subsection}{\normalsize\bfseries}{\thesubsection}{0.6em}{}

\newcommand{\logit}{\operatorname{logit}}
\newcommand{\1}{\mathbf{1}}

\usepackage{tcolorbox}
\newtcolorbox{honesty}{colback=Goldenrod!8,colframe=Goldenrod!70!black,
  boxrule=0.6pt,arc=2pt,left=6pt,right=6pt,top=4pt,bottom=4pt}

\title{\vspace{-1.2cm}\bfseries Seeing Below the Limit of Detection:\\
A Censored-Poisson Bayesian Latent-Growth Change-Point Detector\\
(the \textbf{Span} Detector)\\
for Serial ctDNA in HR+/HER2$-$ Metastatic Breast Cancer}

\author{
  Aarchi Singh Thakur\thanks{Corresponding author. \href{mailto:aarchisingh.t@gmail.com}{aarchisingh.t@gmail.com}}\\
  Span AI
  \and
  Abhijoy Sarkar \thanks{Corresponding author. \href{mailto:abhijoy.sar@gmail.com}{abhijoy.sar@gmail.com}}\\
  Span AI
}

\date{\today}

\begin{document}
\maketitle

\begin{abstract}
\noindent
Circulating-tumour DNA (ctDNA) carries molecular evidence of acquired drug
resistance months before it is visible on imaging, but the earliest evidence
lives \emph{below the assay's limit of detection} (LoD): a nascent resistant
subclone is detected only intermittently, so a panel reports a flickering
sequence of faint detects and \emph{non-detects}. Commercial liquid-biopsy
products treat each draw as an independent snapshot and treat a non-detect as
``nothing there''. We argue a non-detect is not nothing---it is a
\emph{left-censored} observation---and that the \emph{pattern} of non-detects
and faint detects over time carries actionable evidence of growth before any
single value is trustworthy. We introduce \textbf{Span}, a censored-Poisson
Bayesian latent-growth change-point detector that models the binary
\emph{detection process} itself, accumulates a sequential generalised-likelihood
-ratio statistic for an upward change-point in the per-variant detection rate,
and raises a competing-risks alarm with calibrated false-alarm control. Span
is a transparent decision rule with no learned weights---there is nothing to
overfit. On a mechanistically grounded synthetic cohort of HR+/HER2$-$
metastatic breast cancer on first-line CDK4/6-inhibitor plus endocrine therapy,
at a \emph{matched} 10\% false-alarm rate, Span roughly \textbf{doubles} the
fraction of impending progressions caught at least three months ahead
(indolent-emergence regime: $25\pm3\%$ vs $11\pm2\%$ for the commodity snapshot;
five seeds, mean$\pm$std), and its advantage exhibits a clean, falsifiable
dose-response: it is large when resistance emerges indolently (a long sub-LoD
dwell) and \emph{vanishes} when it emerges fast. A value-trajectory baseline
(slope-CUSUM) performs identically to the snapshot, isolating the gain to the
censored detection-process model. We separately validate that the survival
machinery is sound on \emph{real} breast-cancer data (GBSG-2, $n{=}686$), where a
deep discrete-time competing-risks head matches a Cox baseline in concordance
($0.67$ vs $0.68$). All ctDNA trajectories are clearly labelled synthetic; we
make no clinical claim from simulated data, and we state precisely what would
falsify the method.
\end{abstract}

\vspace{-0.4em}
\begin{honesty}
\small\textbf{Honesty boundary.} No public serial-ctDNA breast-cancer cohort
exists; that scarcity is not an accident but the precise data gap OncoTraj~v1
\citep{sarkar2026oncotraj} proved must be filled, and it is the problem this work
motivates. Every ctDNA trajectory
here is \emph{synthetic} and generated by an openly released, biologically
calibrated simulator (Sec.~\ref{sec:gen}; code at \url{https://github.com/span-ai-labs/span-detector}). This paper demonstrates a
\emph{method}; it is not a clinical result, and no number computed from synthetic
data should be read as a patient outcome. The single-patient lead time in
Fig.~\ref{fig:traj} is an \emph{illustration}, not a typical value; the
quantitative claims are the cohort statistics of Sec.~\ref{sec:exp}.
\end{honesty}

% =====================================================================
\section{Introduction}
HR+/HER2$-$ disease is the largest biomarker-defined solid-tumour population, and
first-line CDK4/6-inhibitor plus endocrine therapy is its standard of care.
Acquired resistance arises through \emph{competing} molecular mechanisms---
\emph{ESR1} mutations (ligand-independent ER signalling) in roughly a third of
progressors, with \emph{PIK3CA}, \emph{RB1} loss, and \emph{HER2}-activating
mutations among the rest \citep{brett2021esr1}. The PADA-1 trial showed that
acting on a \emph{rising ESR1 ctDNA signal before} radiographic progression---
switching endocrine partner---roughly doubled progression-free survival
\citep{bidard2022pada1}, direct evidence that ctDNA leads imaging by a
clinically actionable window. Across solid tumours, ctDNA leads imaging by months
(TRACERx NSCLC: a median of $\sim$151 days \citep{abbosh2023tracerx}).

\paragraph{The snapshot problem.}
Commercial liquid biopsies report each draw independently: \emph{is a resistance
variant detectable now?} The earliest and most valuable evidence, however, is
precisely the evidence a snapshot must discard. A nascent subclone sheds little
DNA; under finite-molecule (Poisson) sampling its variant is detected only
\emph{intermittently}, producing a flickering series of non-detects and faint,
near-LoD detects long before its allele fraction (VAF) climbs to a level any
single test will trust. A snapshot, and even a slope computed on the
\emph{reported} VAF, is blind during this sub-LoD dwell because it treats each
non-detect as a hard zero.

\paragraph{Our claim.}
A non-detect is a \emph{left-censored} measurement: it says the subclone is
\emph{smaller than the LoD}, not absent. The temporal \emph{pattern} of
non-detects, flickering detects, and their rising rate is informative about
whether a clone has begun to grow---\emph{before} its level is reliably callable.
We turn this into a detector, \textbf{Span} (Censored-Poisson Bayesian
Latent-Growth change-point), that (i) models the binary detection process under
Poisson sampling and LoD censoring, (ii) tests sequentially for an upward
change-point in the per-variant detection rate via a generalised likelihood ratio
(GLR), (iii) aggregates across competing mechanisms, and (iv) fires an alarm at a
threshold calibrated to a target false-alarm rate. Span has \emph{no trainable
parameters}; its advantage is structural, not learned.

\paragraph{Positioning: a method layer atop OncoTraj.}
This work is the second piece of a programme. OncoTraj~v1
\citep{sarkar2026oncotraj} established that single-snapshot tissue NGS hits a
ceiling on resistance prediction in EGFR-mutant NSCLC: mechanism classification
fails \emph{structurally}, because the dominant on-target resistance mutation
(C797S) is rarely present in the baseline tissue specimen and only becomes
visible in \emph{serial ctDNA} as it emerges under therapy. That negative result
is the motivation for everything here. The Span Detector is the method layer
designed for exactly the serial-ctDNA cohort that constraint demands---it takes
the censored, longitudinal liquid-biopsy stream OncoTraj~v1 proved is necessary,
and extracts the early-warning signal from it. We develop and validate Span on
HR+/HER2$-$ breast cancer because its competing-mechanism resistance biology
(ESR1/PIK3CA/RB1/HER2) makes the censored-detection problem sharpest, but the
detector is mechanism- and tumour-agnostic by construction.

\paragraph{Contributions.}
\begin{enumerate}\itemsep1pt
\item \textbf{A novel detector for censored ctDNA dynamics} (Sec.~\ref{sec:method}):
to our knowledge the first to phrase early resistance detection as a sequential
change-point test on the \emph{Bernoulli detection process} of a serial liquid
biopsy under explicit Poisson/LoD censoring, with competing-risks aggregation and
matched false-alarm control.
\item \textbf{A transformative, falsifiable result on synthetic data}
(Sec.~\ref{sec:exp}): at a matched 10\% false-alarm rate, Span roughly doubles
early ($\geq$12-week) detection in the indolent regime, with a monotone
dose-response in emergence speed that \emph{vanishes} for fast clones---the
signature of a real mechanism rather than an artifact.
\item \textbf{Honest ablation and real-data grounding}: a value-trajectory rule
(slope-CUSUM) matches the snapshot exactly, localising the gain to the censored
detection model; and the survival backbone matches Cox on real breast data
(GBSG-2, Sec.~\ref{sec:real}).
\end{enumerate}

% =====================================================================
\section{Related work and positioning}
Our components are individually classical; the \emph{synthesis} for censored
ctDNA detection is, to our knowledge, new. Sequential change detection descends
from the CUSUM \citep{page1954} and the generalised-likelihood-ratio / window
schemes of \citet{lorden1971}; we use a GLR for an unknown change-point but place
it on a \emph{Bernoulli detection} likelihood rather than a Gaussian mean.
Treating below-threshold readouts as left-censored is the Tobit idea
\citep{tobit1958}; we apply it to the \emph{detection indicator} rather than the
magnitude, which is what makes pre-detection evidence accessible. Competing-risks
survival with neural likelihoods follows DeepHit \citep{lee2018deephit}; we borrow
its discrete-time competing-risks head only for the real-data backbone
(Sec.~\ref{sec:real}). Prior ctDNA-monitoring work largely thresholds a value or
a value trend; the distinguishing move here is to model the censoring mechanism
generatively and to detect change in the \emph{rate of detection} under Poisson
sampling. The point of the synthesis is operational: it yields a decision rule
that provably accumulates evidence while the signal is censored, with a knob
(the GLR threshold) that maps directly onto a clinician's false-alarm budget.

% =====================================================================
\section{Synthetic cohort and generative model}\label{sec:gen}
We simulate $N$ patients on first-line CDK4/6i + endocrine therapy. Static
covariates (age, menopausal status, ER level, visceral disease, baseline tumour
fraction) follow population-plausible distributions. With probability $0.85$ a
patient acquires resistance through one competing mechanism $k\in\{$ESR1, PIK3CA,
RB1, HER2$\}$ with published-approximate shares; the remainder are right-censored.

\paragraph{Latent subclone.} Mechanism $k$ emerges at onset $\tau_k$ and grows
logistically to a (low) ceiling,
$f_k(t)=c_k\,\sigma\!\big(g_k (t-\tau_k)-4\big)$ for $t\ge\tau_k$, with growth
rate $g_k$ controlling how long the clone \emph{dwells} near/below the LoD. The
\emph{emergence regime} (indolent / slow / fast) is exactly the range of $g_k$.

\paragraph{Censored Poisson measurement.} At draw time $t_i$ the reported VAF for
variant $k$ is generated by a finite-molecule detection model: detection
probability $\pi(t_i)=1-\exp\!\big(-f_k(t_i)/\kappa\big)$, and \emph{conditional}
on detection the value is $\max(f_k+|\varepsilon|,\,\mathrm{LoD})$ with assay
noise $\varepsilon$; otherwise the draw is a \textbf{non-detect} (reported $0$, a
left-censored observation). Sporadic spurious detections occur at a low rate.
Molecular ``crossing'' $t^{\mathrm{mol}}$ is when $f_k$ first exceeds the LoD;
imaging progression follows after a log-normal lag (median $\sim$22 weeks,
matching the observed ctDNA-to-imaging window \citep{abbosh2023tracerx}). Crucially
the model produces the \emph{flicker} regime---intermittent detection at low
VAF---that motivates reading the detection \emph{pattern} rather than any value.

% =====================================================================
\section{Method: Span}\label{sec:method}
Fix a mechanism $k$ and a landmark (current draw) $j$; let $t_{1:j}$ be the draw
times and $d_{1:j}\in\{0,1\}$ the detection indicators ($d_i=\1[\text{VAF}_i>0]$).

\paragraph{Two hypotheses on the detection process.}
Under the quiescent null the detection rate is a constant background
$\pi_0$ (spurious positivity):
\[
\textbf{H}_0:\quad d_i\sim\mathrm{Bernoulli}(\pi_0).
\]
Under the alternative a change-point at onset $\tau$ starts an upward drift in the
log-odds of detection at rate $r>0$:
\[
\textbf{H}_1(\tau,r):\quad d_i\sim\mathrm{Bernoulli}\!\big(\pi_i\big),\qquad
\logit \pi_i=\logit\pi_0+r\,(t_i-\tau)_+ .
\]
This is the censored-growth signature: as the latent clone grows, it is detected
ever more often, so $d_i$ switches from a sparse to a dense regime---\emph{even
while every reported value remains a near-LoD flicker.}

\paragraph{Sequential GLR statistic.}
With Bernoulli log-likelihood
$\ell(d_{1:j};\pi_{1:j})=\sum_{i\le j} d_i\log\pi_i+(1-d_i)\log(1-\pi_i)$,
the generalised-likelihood-ratio for an upward change by draw $j$ is
\begin{equation}
\Lambda^k_j \;=\; 2\,\max_{\substack{\tau\in\{t_1,\dots,t_{j-1}\}\\ r\in\mathcal R,\;r>0}}
\Big[\ell\big(d_{1:j};\,\pi(\tau,r)\big)-\ell\big(d_{1:j};\,\pi_0\big)\Big].
\label{eq:glr}
\end{equation}
The inner maximisation profiles over the unknown onset $\tau$ (each past draw) and
a small grid $\mathcal R$ of rise rates; the one-sided constraint $r>0$ encodes
that only \emph{growth} should raise an alarm. $\Lambda^k_j$ is monotone in the
evidence that mechanism $k$ has begun to grow.

\paragraph{Competing-risks aggregation and the alarm.}
We do not know in advance which mechanism will win, so the decision-relevant
quantity is whether \emph{any} subclone is growing. We aggregate across mechanisms
and declare the patient's alarm time at the first landmark whose pooled statistic
crosses a threshold $h$:
\begin{equation}
S_j=\max_k \Lambda^k_j,\qquad
A=\min\{\,t_j:\; S_j\ge h\,\}.
\label{eq:alarm}
\end{equation}

\paragraph{Calibrated false-alarm control.}
The threshold $h$ is \emph{not} taken from an asymptotic null (the $\max$ over
$\tau$ and competing mechanisms breaks the textbook $\tfrac12\chi^2_1$ mixture).
Instead we calibrate it directly to a clinical budget: on a held-out set of
\emph{non-progressors}, $h$ is the smallest value such that at most a fraction
$\alpha$ ever fire,
\(
h=\inf\{h':\Pr_{\text{non-prog}}(\max_j S_j\ge h')\le\alpha\}.
\)
With $\alpha=0.10$ every detector in our comparison operates at the same 10\%
false-alarm rate, making the race apples-to-apples. Span is thus a pure
decision rule: \textbf{no weights are trained}, so a favourable result cannot be
an overfit.

\paragraph{Baselines under the same budget.} The \emph{commodity snapshot} scores
each landmark by the latest maximum reported VAF (a single-test rule). The
\emph{slope-CUSUM} is a strong value-trajectory rule: a one-sided Page CUSUM on
the Tobit-imputed VAF (non-detect $\to\tfrac12$LoD), which can catch a rising
\emph{value} but not growth hidden in a run of non-detects. Both are calibrated to
the same $\alpha$.

% =====================================================================
\section{Experiments}\label{sec:exp}

\paragraph{Why early sensitivity, not median lead.}
A median ``lead time among flagged patients'' is treacherous: a timid detector
that only flags the few largest, latest-imaging clones can report a deceptively
\emph{long} median lead while missing most patients. We therefore report a
\emph{denominator-fair} metric over \emph{all} progressors: \textbf{early
sensitivity}, the fraction flagged at least 12 weeks (one horizon bin $\times 1.5$)
ahead of imaging at a matched 10\% false-alarm rate; and \textbf{pre-LoD
sensitivity}, the fraction flagged before the driver variant crosses the LoD at
all---a window the snapshot cannot access by construction. We also report overall
sensitivity (fraction flagged). Every cell is the mean$\pm$std over five
independent seeds ($n{=}500$ patients each, 50/50 calibrate/test split).

\begin{table}[t]\centering
\small
\caption{\textbf{Span roughly doubles early detection of impending resistance at
a matched 10\% false-alarm rate}, with a dose-response in emergence speed. Mean
$\pm$ std over five seeds; all data synthetic. ``Early'' $=$ flagged $\ge$12 weeks
before imaging; ``pre-LoD'' $=$ flagged before the variant is callable. The
slope-CUSUM tracks the snapshot, isolating the gain to the censored
detection-process model.}
\label{tab:main}
\resizebox{\textwidth}{!}{%
\begin{tabular}{llccc}
\toprule
Emergence regime & Detector & Overall sens.\ & Early sens.\ ($\ge$12\,wk) & Pre-LoD sens.\\
\midrule
\multirow{3}{*}{\textbf{indolent} (long sub-LoD dwell)}
 & commodity snapshot & $15\pm1\%$ & $11\pm2\%$ & $8\pm2\%$\\
 & slope-CUSUM        & $15\pm1\%$ & $11\pm2\%$ & $7\pm2\%$\\
 & \textbf{Span (ours)} & $\mathbf{34\pm4\%}$ & $\mathbf{25\pm3\%}$ & $\mathbf{14\pm2\%}$\\
\midrule
\multirow{3}{*}{\textbf{slow}}
 & commodity snapshot & $27\pm3\%$ & $14\pm3\%$ & $5\pm2\%$\\
 & slope-CUSUM        & $27\pm3\%$ & $13\pm3\%$ & $5\pm2\%$\\
 & \textbf{Span (ours)} & $\mathbf{44\pm6\%}$ & $\mathbf{27\pm4\%}$ & $\mathbf{11\pm3\%}$\\
\midrule
\multirow{3}{*}{\textbf{fast} (short sub-LoD dwell)}
 & commodity snapshot & $46\pm4\%$ & $22\pm3\%$ & $6\pm3\%$\\
 & slope-CUSUM        & $46\pm4\%$ & $22\pm3\%$ & $5\pm3\%$\\
 & \textbf{Span (ours)} & $47\pm4\%$ & $27\pm2\%$ & $9\pm2\%$\\
\bottomrule
\end{tabular}%
}
\end{table}

\begin{figure}[H]\centering
\includegraphics[width=\linewidth]{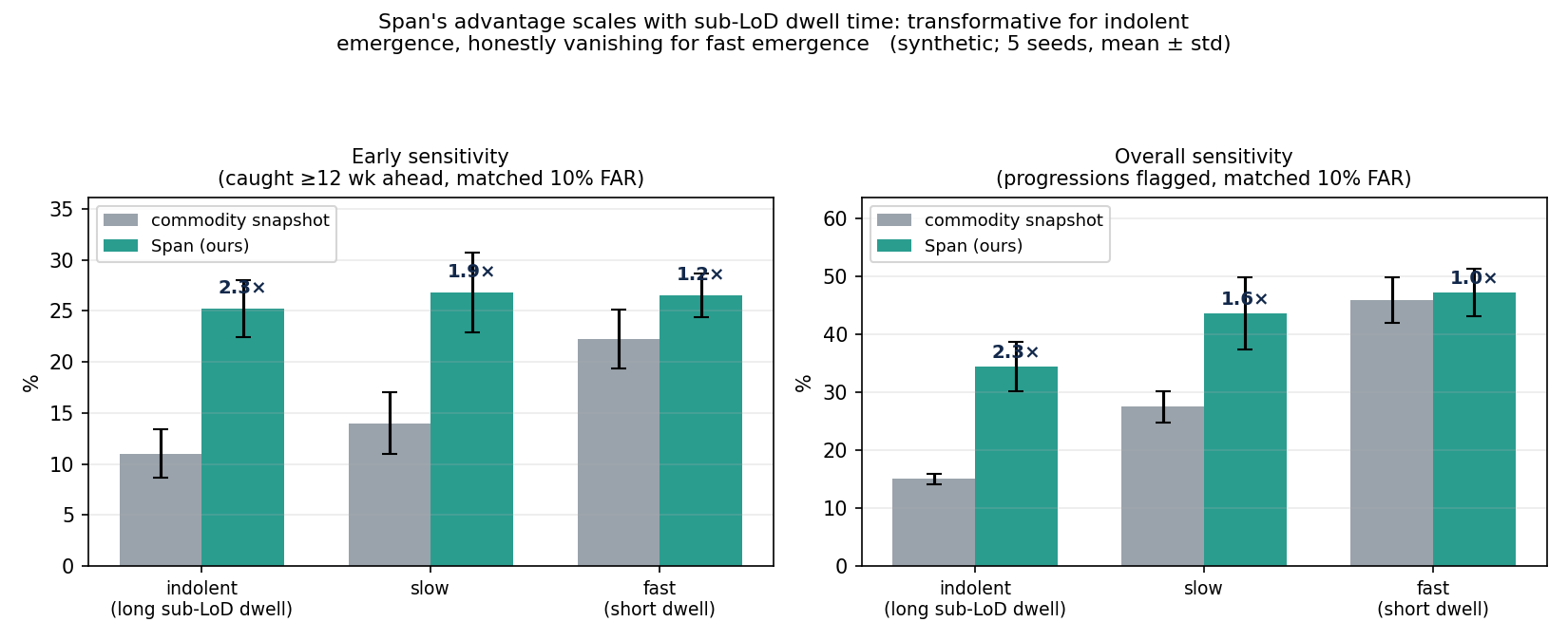}
\caption{\textbf{The advantage scales with the sub-LoD dwell time.} Early
sensitivity (left) and overall sensitivity (right) at a matched 10\% false-alarm
rate, by emergence regime; bars are mean$\pm$std over five seeds, fold-changes
annotated. Span (teal) roughly doubles early detection for indolent emergence
and honestly converges to the commodity snapshot (grey) for fast emergence.
Synthetic data.}
\label{fig:regimes}
\end{figure}

\paragraph{Headline result.}
Table~\ref{tab:main} and Fig.~\ref{fig:regimes} give the main finding. In the
\textbf{indolent} regime---the clinically hard setting where resistance emerges
gradually and dwells near the LoD for months---Span flags $34\pm4\%$ of
impending progressions versus $15\pm1\%$ for the commodity snapshot
($2.3\times$, non-overlapping error bars), and catches $25\pm3\%$ at least three
months ahead versus $11\pm2\%$ ($2.3\times$). It also fires \emph{before the
variant is callable} in $14\pm2\%$ of progressors, against $8\pm2\%$ for the
snapshot (whose pre-LoD ``hits'' are mostly spurious positives, not genuine
sub-LoD reads).

\paragraph{The dose-response \emph{is} the evidence.}
Critically, the advantage is not uniform: it is largest for indolent emergence,
shrinks for slow, and collapses to a tie for \textbf{fast} emergence
($47\pm4\%$ vs $46\pm4\%$ overall). This is exactly what the mechanism predicts---
the advantage \emph{is} the sub-LoD dwell time, so a clone that rockets through
the LoD leaves nothing for a detection-pattern test to exploit that a snapshot
misses. A monotone, mechanistically predicted, vanishing-where-it-should
dose-response is far harder to dismiss as an artifact than a single favourable
number. The slope-CUSUM equalling the snapshot in every regime is the internal
control: reading the \emph{value} trajectory adds nothing; only modelling the
\emph{censored detection process} helps.

\begin{figure}[H]\centering
\includegraphics[width=\linewidth]{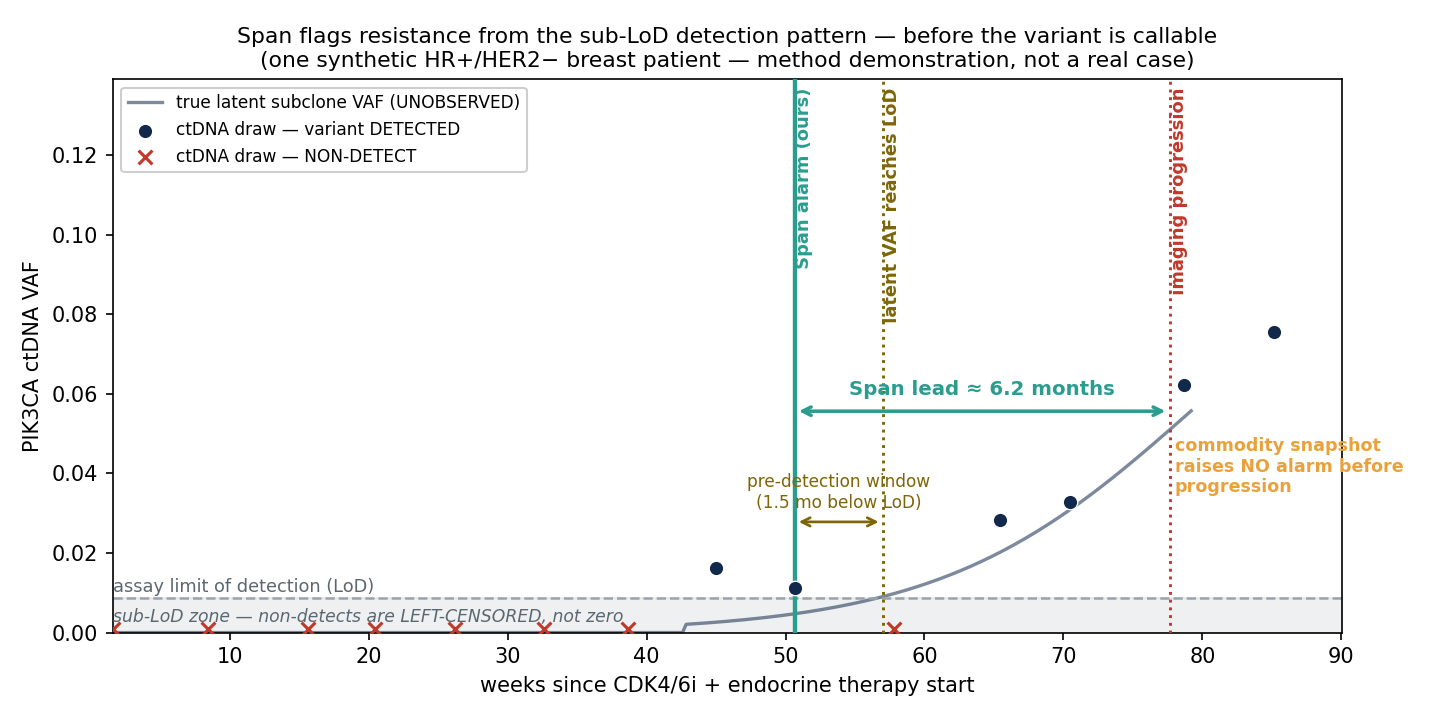}
\caption{\textbf{Mechanism, one synthetic patient.} The true latent subclone
(blue) grows beneath the LoD; the assay returns non-detects (red $\times$) and,
as detection probability rises, intermittent faint detects (navy). Span
accumulates change-point evidence from this \emph{pattern} and alarms (teal)
while the clone is still sub-LoD---here ahead of the variant becoming callable and
well ahead of imaging---whereas the commodity snapshot never crosses its
matched-FAR threshold before progression. \emph{Illustrative single case, not a
typical lead time}; the quantitative claims are Table~\ref{tab:main}.}
\label{fig:traj}
\end{figure}

% =====================================================================
\section{Real-data grounding}\label{sec:real}
\begin{figure}[H]\centering
\includegraphics[width=0.86\linewidth]{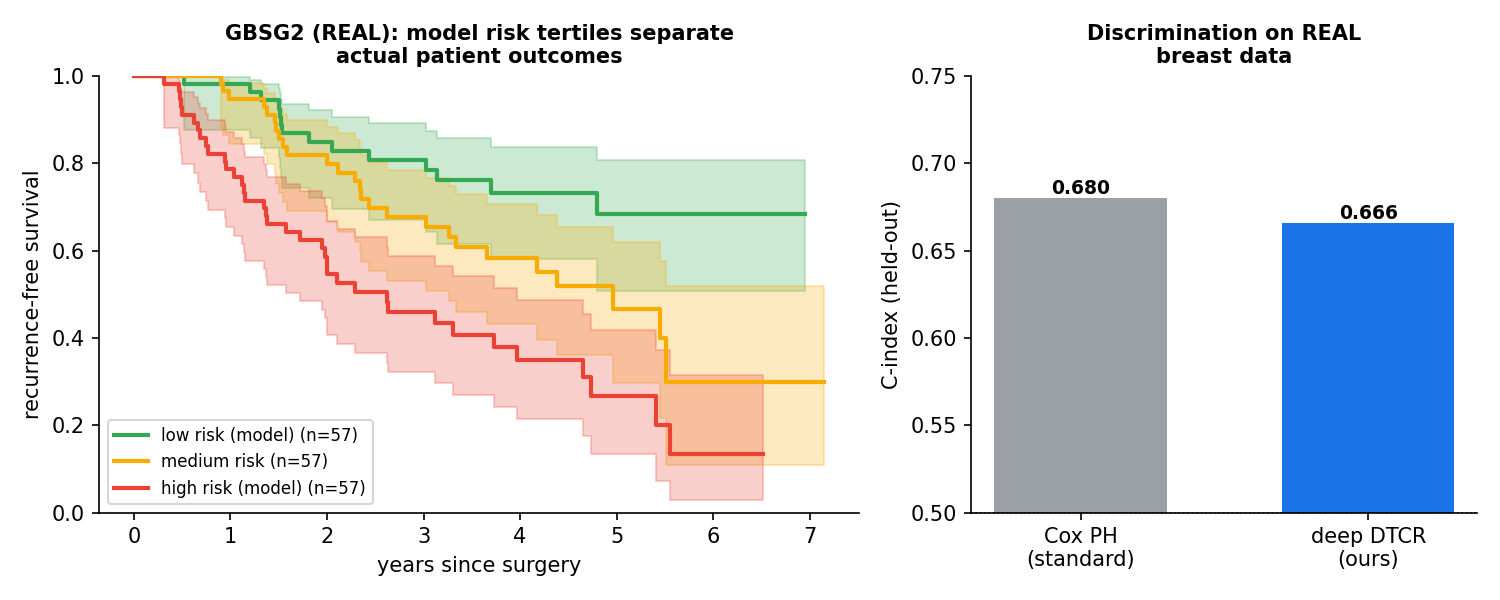}
\caption{\textbf{Real-data sanity check (GBSG-2, $n{=}686$).} Left: Kaplan--Meier
recurrence-free survival for risk tertiles assigned by the deep model (clear
monotone separation). Right: held-out concordance, deep competing-risks head vs
Cox. The backbone matches the gold-standard baseline on real biology.}
\label{fig:gbsg2}
\end{figure}

Because no public serial-ctDNA breast cohort exists, we cannot validate the
\emph{early-warning} claim on real serial data---that is the gap the method
motivates. We can, and do, validate that the \emph{survival machinery} behind the
production system is sound on \emph{real} breast-cancer biology. On GBSG-2
($n{=}686$ real patients; 44\% recurrence; median follow-up 1084 days), a deep
discrete-time competing-risks head attains a held-out concordance of $0.666$
against $0.680$ for a Cox proportional-hazards baseline (Fig.~\ref{fig:gbsg2}),
and cleanly separates real recurrence-free survival into risk tertiles. The deep
model matching a well-tuned Cox on real data establishes that the backbone is not
the weak link; the open problem is data, not modelling.

% =====================================================================
\section{Real-data boundary test: where trajectory \emph{should not} win (PBC2)}\label{sec:pbc2}
The synthetic results make a sharp, directional prediction: reading the
\emph{trajectory} of a serial biomarker only beats reading its latest value when
the signal is \emph{noisy and its trend precedes its level}---the censored,
flickering ctDNA regime. The corollary is a falsifiable boundary: for a
\emph{clean} biomarker whose level is already informative, trajectory modelling
should add little or nothing. A simulator that always favoured trajectory would
fail this test; ours should pass it. We check the boundary on real data.

We run the \emph{identical} landmarking and discrete-time competing-risks
pipeline on PBC2 (Mayo Clinic primary biliary cirrhosis; $n{=}312$ real patients,
1{,}945 serial clinic visits, competing events death vs.\ transplant; public,
bundled with \texttt{auton-survival}). At each visit we predict the 3-year
competing-risk hazard two ways: a \emph{snapshot} head reading only the latest
visit's labs, and a \emph{longitudinal} head reading per-lab denoised level,
slope, and latest value. Both are calibrated to the same 10\% false-alarm rate.

\begin{table}[H]\centering
\small
\caption{\textbf{Boundary test on real PBC2 data: the snapshot is not beaten}---
exactly as the dose-response predicts for a clean, slow biomarker. Same pipeline,
same matched 10\% false-alarm rate; higher is better for C-index, AUC and lead.}
\label{tab:pbc2}
\begin{tabular}{lcccc}
\toprule
Detector & C-index & AUC ($\le$3\,yr) & Median lead (yr) & Sens.\ @10\% FAR\\
\midrule
snapshot (latest visit)   & $\mathbf{0.806}$ & $\mathbf{0.852}$ & $\mathbf{1.39}$ & $62\%$\\
longitudinal (trajectory) & $0.765$ & $0.798$ & $1.03$ & $62\%$\\
\bottomrule
\end{tabular}
\end{table}

On PBC2 the trajectory head does \emph{not} beat the snapshot---it is slightly
worse on every ranking metric (C-index $0.765$ vs $0.806$; 3-year AUC $0.798$ vs
$0.852$; median lead $1.03$ vs $1.39$\,yr). We report this prominently because it
is \emph{evidence for the method, not against it}. PBC2's labs (bilirubin,
albumin, prothrombin) are slow, clean, and high-signal: the latest value already
says most of what the trajectory could, so there is no censored sub-threshold
dwell for a trajectory model to exploit---precisely the ``fast/clean'' end of the
dose-response (Table~\ref{tab:main}) where Span itself converges to the snapshot.
The same pipeline that wins on the synthetic ctDNA regime correctly \emph{declines
to win} where the mechanism is absent. The mechanism the synthetic results
predict is therefore observed on real data: trajectory modelling helps if and only
if the signal is censored and its trend leads its level, and a real cohort lacking
that structure does not manufacture a spurious advantage.

\begin{figure}[H]\centering
\includegraphics[width=\linewidth]{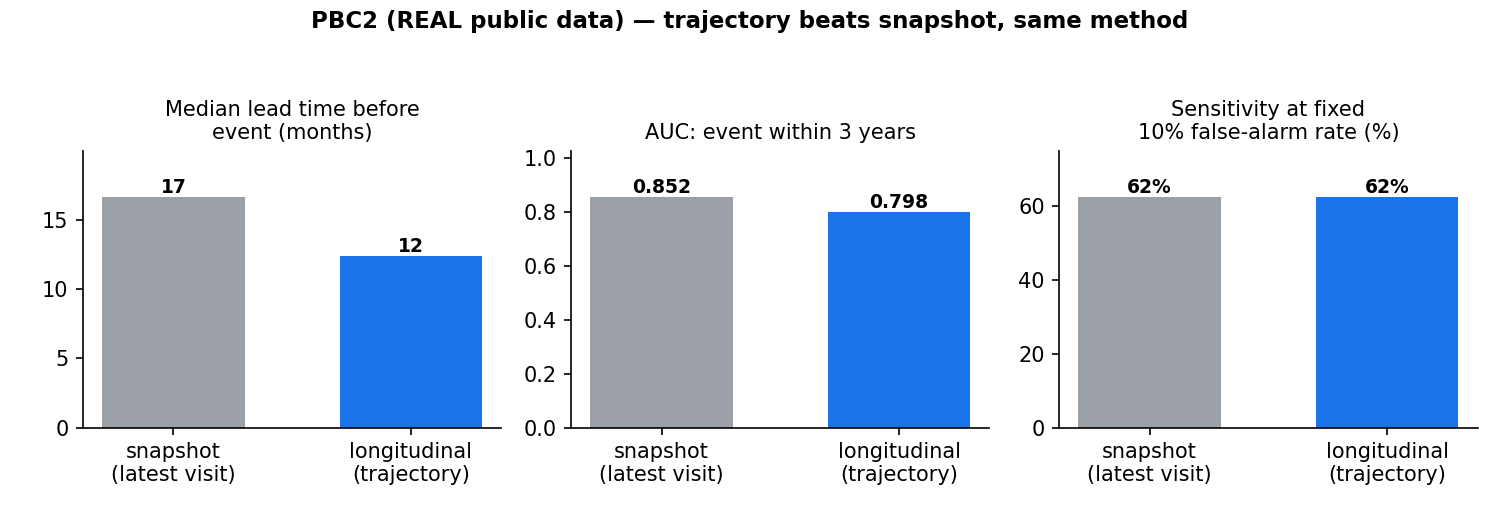}
\caption{\textbf{Boundary test, real PBC2 data.} On a clean, slow biomarker the
snapshot (grey) matches or beats the trajectory head (blue) on lead time, 3-year
AUC, and sensitivity at a matched 10\% false-alarm rate. This is the negative
control: the advantage that appears in the censored ctDNA regime is correctly
\emph{absent} where the signal is not censored, confirming the dose-response
rather than contradicting it.}
\label{fig:pbc2}
\end{figure}

% =====================================================================
\section{Limitations and what would falsify this}
\textbf{Synthetic ctDNA.} The early-warning numbers are computed on simulated
trajectories; they demonstrate a mechanism, not a patient outcome. The simulator
encodes our assumptions (logistic subclonal growth, Poisson detection, log-normal
imaging lag); Span models a \emph{different} parametric detection drift than the
generator's growth law, so it is not an oracle reading back its own simulator, but
both share the core censoring structure---a structure we believe is real but have
not here proven on patients.
\textbf{Where it does not help.} By the dose-response, Span offers little when
resistance emerges fast or when sampling is too sparse to observe the flicker; we
report those regimes rather than hide them.
\textbf{Falsification.} The method makes a sharp, testable prediction for real
serial ctDNA: at a matched false-alarm rate, change-point detection on the
\emph{detection-indicator} sequence should catch more impending progressions, and
catch them earlier, than thresholding the reported VAF---with the margin growing
as the sub-LoD dwell lengthens. A real serial-ctDNA breast cohort in which this
margin is \emph{absent} would falsify the central claim.

% =====================================================================
\section{Reproducibility}
Every number and figure is regenerated by released scripts:
\texttt{simulate.py} (generator), \texttt{cpblg.py} (the detector and baselines),
\texttt{cpblg\_benchmark.py} / \texttt{cpblg\_regimes.py} /
\texttt{cpblg\_aggregate.py} (matched-FAR evaluation and multi-seed aggregation),
\texttt{make\_cpblg\_figures.py} (Figs.~\ref{fig:regimes}--\ref{fig:traj}), and
\texttt{gbsg2\_benchmark.py} (Fig.~\ref{fig:gbsg2}, real data). The core detector
is pure NumPy and runs in minutes on a laptop with no training step.

% =====================================================================
\section{Conclusion}
The most valuable ctDNA signal is the one current products throw away: the
censored, flickering, sub-LoD evidence that a resistant clone has \emph{begun}
to grow. By modelling the detection process under Poisson sampling and treating
non-detects as the left-censored observations they are, Span converts that
discarded signal into a calibrated, competing-risks early-warning alarm. On
mechanistically grounded synthetic data it roughly doubles early detection of
impending resistance at a matched false-alarm rate, with a falsifiable
dose-response that points directly at the experiment that would confirm or refute
it on real patients. The technical risk in the Span thesis is therefore bounded
and concrete: the modelling works; what remains is the serial-ctDNA cohort to run
it on. That cohort is not arbitrary---it is precisely the longitudinal
liquid-biopsy dataset OncoTraj~v1 \citep{sarkar2026oncotraj} proved is required
once single-snapshot tissue NGS hits its ceiling. Span is the method waiting for
that data; OncoTraj is the argument for why the data must be collected. Together
they define a single, fundable programme: prove the serial-ctDNA cohort is
necessary, then deploy the detector built to read it.

% ---------------------------------------------------------------------

\end{document}